\documentclass{emulateapj}
\usepackage{apjfonts}
\usepackage{epsfig,color}
\usepackage{mathrsfs}
\usepackage{ulem}
\usepackage{amsmath}
\usepackage{empheq}
\usepackage{bm}

\def\bhm{M_{\bullet}}

\def\ergs{{\rm erg~s^{-1}}}
\def\fblr{f_{_{\rm BLR}}}

\def\kms{\rm km~s^{-1}}

\def\LEdd{L_{\rm Edd}}

\def\mathdotM{\dot{\mathscr{M}}}

\def\RFe{{\cal R}_{\rm Fe}}
\def\rhb{R_{_{\rm H\beta}}}

\def\sighb{\sigma_{_{\rm H\beta}}}
\def\sunm{M_{\odot}}
\def\tauhb{\tau_{_{\rm H\beta}}}

\def\Dhb{{\cal D}_{_{\rm H\beta}}}

\def\calA{{\cal A}}

\def\feii{Fe {\sc ii}}

\def\Lbol{L_{\rm bol}}

\journalinfo{To appear in {\it The Astrophysical Journal Letters}.}
\slugcomment{Received 2015 November 21;  accepted 2015 December 30}

\begin{document}

\title{The Fundamental Plane of the Broad-line Region in Active Galactic Nuclei}

\author{
Pu Du\altaffilmark{1},
Jian-Min Wang\altaffilmark{1,2,*},
Chen Hu\altaffilmark{1},
Luis C. Ho\altaffilmark{3,4},
Yan-Rong Li\altaffilmark{1} and
Jin-Ming Bai\altaffilmark{5}
}

\altaffiltext{1}
{Key Laboratory for Particle Astrophysics, Institute of High Energy Physics,
Chinese Academy of Sciences, 19B Yuquan Road, Beijing 100049, China}

\altaffiltext{2}
{National Astronomical Observatories of China, Chinese Academy of Sciences,
 20A Datun Road, Beijing 100020, China}

\altaffiltext{3}
{Kavli Institute for Astronomy and Astrophysics, Peking University, Beijing 100871, China} 

\altaffiltext{4}
{Department of Astronomy, School of Physics, Peking University, Beijing 100871, China} 

\altaffiltext{5}{Yunnan Observatories, Chinese Academy of Sciences, Kunming 650011, China}
\altaffiltext{*}{Corresponding author, wangjm@mail.ihep.ac.cn}

\begin{abstract}
Broad emission lines in active galactic nuclei (AGNs) mainly arise 
from gas photoionized by continuum radiation from an accretion disk around a 
central black hole. The shape of the broad-line profile, described by 
$\Dhb={\rm FWHM}/\sighb$, the ratio of full width at half maximum to the 
dispersion of broad H$\beta$, reflects the dynamics of the broad-line region 
(BLR) and correlates with the dimensionless accretion rate ($\mathdotM$) or 
Eddington ratio ($\Lbol/\LEdd$).  At the same time, $\mathdotM$ and 
$\Lbol/\LEdd$ correlate with $\RFe$, the ratio of optical \feii\, to H$\beta$ 
line flux emission. Assembling all AGNs with reverberation mapping measurements 
of broad H$\beta$, both from the literature and from new observations reported 
here, we find a strong bivariate correlation of the form 
$\log(\mathdotM,L_{\rm bol}/L_{\rm Edd})=\alpha+\beta\Dhb+\gamma\RFe,$
where $\alpha=(2.47,0.31)$, $\beta=-(1.59,0.82)$ and $\gamma=(1.34,0.80)$. We 
refer to this as the fundamental plane of the BLR.  We apply the plane to a 
sample of $z < 0.8$ quasars to demonstrate the prevalence of super-Eddington 
accreting AGNs are quite common at low redshifts.
\end{abstract}


\section{Introduction}
Broad emission lines are a hallmark feature of type 1 active galactic nuclei 
(AGNs) and quasars \citep{osterbrock1986}.  As pervasive as they are, 
many basic properties of the broad-line region (BLR), such as its basic 
geometry, dynamics, and physical connection to the accretion disk around the 
supermassive black hole (BH), remain ill-defined.  AGN spectra exhibit both 
tremendous diversity as well as discernable patterns of systematic 
regularity.  Principal component analysis has isolated several dominant 
relationships among emission-line properties \citep{boroson1992, sulentic2000}. 
The main varying trend of those properties, which is so-called Eigenvector 
1 (EV1), has been demonstrated to be driven by Eddington ratios, 
$L_{\rm bol}/L_{\rm Edd}$, where $L_{\rm bol}$ is the bolometric luminosity and 
the Eddington luminosity $L_{\rm Edd} = 1.5 \times 10^{38}\, (\bhm/M_{\odot})$ 
\citep{boroson1992, sulentic2000, shen2014}. As one of the most prominent 
variables in EV1, the relative strength of broad optical \feii\ emission, expressed as 

\begin{equation}
\RFe=\frac{F_{\rm FeII}}{F_{\rm H\beta}}, 
\end{equation}
may correlate with $L_{\rm bol}/L_{\rm Edd}$. 
Sources with high Eddington ratios (accretion 
rates), for instance so-called narrow-line Seyfert 1 galaxies \citep{osterbrock1985}, 
emit exceptionally strong \feii\ lines compared with the normal ones
\citep{boroson1992, hu2008, dong2011}. 
However, the underlying physical mechanism that controls $\RFe$ remains unclear, 
as the formation of \feii\ is very complex (e.g., \citealt{baldwin2004}). 
It may be influenced by different hydrogen density of BLR gas \citep{verner2004}, 
or diverse contribution from microturbulence \citep{baldwin2004}.
In addition, \feii\, lags are generally longer by a factor of a few than H$\beta$ 
in broad-line Seyfert 1 galaxies \citep{barth2013, chelouche2014} and roughly
equal to H$\beta$ lags in narrow-line Seyfert 1s \citep{hu2015},  
implying the potential connection of $\RFe$ with the distribution or structure of 
line-emitting gas. The $\RFe-L_{\rm bol}/L_{\rm Edd}$ correlation indicates that 
Eddington ratios probably regulate all above mentioned properties of BLR. It should 
be noted that $\RFe$ also correlates with some other properties like 
X-ray spectral slopes (e.g., \citealt{wang1996, laor1997, sulentic2000}), 
but it likely originates from relation of those properties and Eddington ratios 
(e.g., \citealt{wang2004, risaliti2009, shemmer2006, brightman2013}).

The overall breadth of the broad emission lines, notably H$\beta$, reflects 
both the virial velocity and inclination of the BLR \citep{kollatschny2011, shen2014}. 
The shape of the line profile may encode more 
information on the detailed dynamics of the BLR (e.g., \citealt{collin2006, kollatschny2011}), 
which itself may depend on fundamental properties 
such as the accretion or outflow rate.  The broad H$\beta$ lines of NLS1s 
tend to have more sharply peaked ($\sim$Lorentzian) profiles compared to 
type 1 AGNs with more normal Eddington ratios \citep{veron2001, zamfir2010}. 
As a non-parametric description of the line profile, one can define
\begin{equation}
\Dhb=\frac{\rm FWHM}{\sighb},
\end{equation}
where $\sighb$ is the dispersion (second moment) of the H$\beta$ line. 
The value of $\Dhb$ is 2.35, 3.46, 2.45, 2.83 and 0 for a Gaussian, a 
rectangular, a triangular, an edge-on rotating ring, and a Lorentzian 
profiles (for a pure Lorentzian profile $\sighb\rightarrow \infty$ and thus 
$\Dhb=0$), respectively (e.g., \citealt{collin2006}). The quantity $\Dhb$
correlates loosely with Eddington ratio \citep{collin2006} and, as the ratio 
of the rotational and turbulent components of the line-emitting clouds 
\citep{kollatschny2011}, gives a simple, convenient parameter that may be 
related to the dynamics of the BLR.

While $\RFe$ and $\Dhb$ each correlates separately with Eddington ratio, we 
demonstrate that both $\RFe$ and $\Dhb$ {\it combined}\ correlate even 
more tightly with Eddington ratio (and dimensionless accretion rate).  This 
bivariate relation, which we call the ``fundamental plane''\footnote{Borrowing 
the terminology from galaxy formation (e.g., \citealt{djorgovski1987}) and 
accreting BHs (e.g., \citealt{merloni2003})} of the BLR links two direct 
observables, plausibly related to the structure and dynamics of the BLR, with 
the dimensionless accretion rate.  Applying the BLR fundamental plane to a 
large sample of Sloan Digital Sky Survey (SDSS) quasars, we find that a 
large fraction of quasars at $z < 0.8$ have super-Eddington accretion rates.

\renewcommand{\arraystretch}{1.2}
\begin{deluxetable*}{lcccccccl}[!t]
\tablecolumns{9}
\setlength{\tabcolsep}{3pt}
\tablewidth{0pc}
\tablecaption{The Sample of Reverberation-mapped AGNs}
\tabletypesize{\footnotesize}
\tablehead{
   \colhead{Objects}                       &
   \colhead{$\log L_{5100}$}               &
   \colhead{$\log\left(\bhm/\sunm\right)$} &
   \colhead{$\log\mathdotM$}               &
   \colhead{FWHM}                          &
   \colhead{$\sigma_{\rm line}$}           &
   \colhead{$\Dhb$}                    &
   \colhead{$\cal{R}_{\rm Fe}$}            &
   \colhead{Ref.}                          \\ \cline{2-9}
   \colhead{}                              &
   \colhead{($\ergs)$}                     &
   \colhead{}                              &
   \colhead{}                              &
   \colhead{($\kms$)}                      &
   \colhead{($\kms$)}                      &
   \colhead{}                              &
   \colhead{}                              &
   \colhead{}
}
\startdata
Mrk 335     & $    43.69\pm0.06 $ & $    6.87_{-0.14}^{+0.10} $ & $     1.17_{-0.30}^{+0.31} $ & $2096\pm170$ & $1470\pm 50$ & $    1.43\pm0.13 $ & $    0.39 $ & 1, 2, 3, 4 \\
           & $    43.76\pm0.06 $ & $    7.02_{-0.12}^{+0.11} $ & $     1.28_{-0.29}^{+0.30} $ & $1792\pm  3$ & $1380\pm 6$ & $    1.30\pm0.01 $ & $    0.77 $ & 4, 5, 6$^a$ \\
           & $    43.84\pm0.06 $ & $    6.84_{-0.25}^{+0.18} $ & $     1.39_{-0.29}^{+0.30} $ & $1679\pm  2$ & $1371\pm 8$ & $    1.23\pm0.01 $ & $    0.77 $ & 4, 5, 6$^a$ \\
           & $    43.74\pm0.06 $ & $    6.92_{-0.14}^{+0.11} $ & $     1.25_{-0.29}^{+0.30} $ & $1724\pm236$ & $1542\pm 66$ & $    1.12\pm0.16 $ & $    0.69 $ & 4, 7$^a$ \\
           & $\bm{43.76\pm0.07}$ & $\bm{6.93_{-0.11}^{+0.10}}$ & $\bm{ 1.27_{-0.17}^{+0.18}}$ & ...          & ...          & $\bm{1.27\pm0.05}$ & $\bm{0.62}$ & 4 \\
PG 0026+129 & $    44.97\pm0.02 $ & $    8.15_{-0.13}^{+0.09} $ & $     0.65_{-0.20}^{+0.28} $ & $2544\pm 56$ & $1738\pm100$ & $    1.46\pm0.09 $ & $    0.33 $ & 4, 5, 8$^a$ \\
PG 0052+251 & $    44.81\pm0.03 $ & $    8.64_{-0.14}^{+0.11} $ & $    -0.59_{-0.25}^{+0.31} $ & $5008\pm 73$ & $2167\pm 30$ & $    2.31\pm0.05 $ & $    0.12 $ & 4, 5, 8$^a$ \\
\enddata
\vspace{-0.2cm}
\tablecomments{All the values of 
 $\log L_{5100}$, $\log (\bhm/\sunm)$ and $\log \mathscr{\dot{M}}$ are compiled from Du et al. (2015).
Values in boldface are the weighted averages of all the measurements for this object. \\ \vglue 0.15cm
\vspace{-0.30cm}
\hspace{0.1in}
Ref.:
 (1) \citealt{du2014}; 
 (2) \citealt{wang2014}; 
 (3) \citealt{hu2015}; 
 (4) \citealt{du2015}; 
 (5) \citealt{collin2006}; 
 (6) \citealt{peterson1998}; 
 (7) \citealt{grier2012}; 
 (8) \citealt{kaspi2000}. 
 \\ \vglue 0.02cm
\vspace{-0.2cm}
 \hspace{0.1in} 
The superscript $a$ for references indicates that $\RFe$ is measured in this 
paper; $b$ indicates that FWHM and $\sighb$ are measured from SDSS spectra (the 
H$\beta$ width of SEAMBHs is significantly broadened by the 5$^{\prime\prime}$ 
longslit of our campaign; see details in Ref. 4); $c$ means the MCMC BH mass 
is used (see Section 2.2); $d$ means that $\Dhb$ is taken from the latest 
measurements in \cite{kollatschny2011}. NGC 5548 marked with $e$ is measured from its mean 
annual spectra in the AGN watch database; the average value is provided here.  
We first calculate $\Dhb$ for each measurement, and then average. In the 
main text, we use these averaged numbers for the objects with multiple RM 
measurements (treated as one point in all figures). For NGC 7469, which was
mapped twice \cite{collier1998} and \cite{peterson2014}, the H$\beta$ lags are not very 
different but the H$\beta$ FWHM is very different; take the values of FWHM 
measured by \cite{kollatschny2011}.  NGC 4051 and PG 1700+518 have very small values of 
$\Dhb$ in Ref. 5, but \cite{kollatschny2011} provides new measurements, which are used 
here.\\ \vglue 0.02cm
\vspace{-0.2cm}
\hspace{0.1in}  This table is available in its entirety in a machine-readable 
form in the online journal. A portion is shown here for guidance regarding its 
form and content.
}
\end{deluxetable*}

\section{Measurements}
\subsection{The Reverberation-mapped AGN sample}
We select all AGNs with reverberation mapping (RM) data (here only broad H$\beta$ 
line), which yield robust BH 
mass estimates needed for our analysis.  All RM AGNs before 2013 are summarized
by \cite{bentz2013}. We took all of 41 AGNs from \cite{bentz2013}.  
Three additional sources (Mrk 1511, NGC 5273, 
KA1858+4850) were subsequently published.  Our project to search for 
super-Eddington accreting massive black holes (SEAMBHs) has monitored about 25 
candidates and successfully measured H$\beta$ lags ($\tauhb$) in 14 AGNs to 
date \citep{du2015} and other five objects monitored between 2014-2015
(to be submitted).  We measure \feii\, using the same approach as \cite{hu2008}
and \cite{hu2015}.
For reverberation-mapped AGNs without published measurements of \feii\ and 
H$\beta$ flux, we fit the mean spectra from the monitoring campaigns, using 
the fitting scheme described in 
\cite{hu2015}. In short, the spectrum is fitted with several components simultaneously: 
(1) a power law for continuum, (2) Fe {\sc ii} template from \cite{boroson1992}, 
(3) host galaxy template if necessary, (4) broad H$\beta$, (5) broad He {\sc ii} 
$\lambda 4686$ emission line, and (6) several Gaussians for narrow lines such as [O {\sc iii}] 
$\lambda\lambda$4959, 5007.  The 
flux of broad optical Fe {\sc ii} is measured by integration from 4434 \AA\ to 4684 \AA.
Table 1 lists the 63 RM AGNs we consider, along with 
the BH mass, 5100 \AA\, luminosity, dimensionless accretion rate, FWHM, 
$\sighb$, $\RFe$ and data sources. 

The sample covers a wide range of accretion rates, $\mathdotM \approx 10^{-3} 
- 10^3$, from the regime of a \cite{shakura1973} standard disk to a 
slim disk \citep{abramowicz1988}.  We take $\RFe$ from the published 
literature if available; otherwise, we measure it from the averaged spectra 
following the spectral fitting scheme of \cite{hu2008, hu2015}.  As the 
variability of H$\beta$ is unusually much larger than that of \feii\ in 
sub-Eddington AGNs, the uncertainties of $\RFe$ are mainly governed by 
H$\beta$ variability, which on average is $\sim 20$\%.  
 
We estimate the BH mass as $\bhm=\fblr V_{\rm FWHM}^2 c\tauhb/G$, where 
$\fblr$ is the virial factor, $V_{\rm FWHM}$ is H$\beta$ FWHM, and $G$ is the 
gravitational constant.  In practice, the factor $\fblr$ is calibrated against
the $\bhm-\sigma$ relation of inactive galaxies \citep{onken2004, ho2014}. 
For consistency with our earlier series of papers, we adopt $\fblr=1$.

\subsection{Accretion rates and Eddington ratios}
We derived accretion rates from the disk model of \cite{shakura1973},
which has been extensively applied to fit the spectra of quasars and Seyfert 1 
galaxies \citep{czerny1987, sun1989, laor1989, collin2002, brocksopp2006, kishimoto2008, 
davis2011, capellupo2015}. The effective temperature distribution is 
given by $T_{\rm eff}=6.2\times 10^{4}\,\dot{m}_{\bullet,0.1}^{1/4}m_7^{1/4}
R_{14}^{-3/4}$\,K, where $\dot{m}_{\bullet,0.1}=\dot{M}_{\bullet}/0.1\sunm\,
{\rm yr^{-1}}$, $\dot{M}_{\bullet}$ is mass accretion rates, $m_7=\bhm/10^7
\sunm$, and $R_{14}=R/10^{14}$cm \citep{frank2002}.  Here the effect of the 
inner boundary is neglected because the region emitting optical radiation is 
far from the boundary. Introducing $x=h\nu/kT_{\rm eff}$, we have the spectral 
luminosity by integrating over the entire disk,
\begin{equation}
L_{\nu}=1.58\times 10^{28}\,\dot{m}_{\bullet,0.1}^{2/3}m_7^{2/3}\nu_{14}^{1/3}\cos i 
        \int_{x_{\rm in}}^{\infty}\frac{x^{5/3}}{e^x-1}dx\,\,\,\,{\rm erg\,s^{-1}\,Hz^{-1}},
\end{equation}
where $i$ is the disk inclination relative to the observer and 
$\nu_{14}=\nu/10^{14}$Hz. Since long-wavelength photons are radiated from 
large disk radii, the integral term in Equation (3) can be well approximated 
by 1.93 for $x_{\rm in}=0$ \citep{davis2011}. We thus have
$\dot{M}_{\bullet}=0.53\left(\ell_{44}/\cos i\right)^{3/2}m_7^{-1}~\sunm~{\rm yr^{-1}}$,
and the dimensionless accretion rate\footnote{The applicability of Eq. (4) 
to SEAMBHs can be justified by the self-similar solution of slim disks \citep{wang1999a, wang1999b}. 
The solution shows that the 5100 \AA\,
photons are emitted from $R_{5100}/R_{\rm Sch}\approx 4.3\times 10^3
m_7^{-1/2}$, and the photon trapping radius $R_{\rm trap}/R_{\rm Sch}\approx 
144\mathdotM_{100}$, where $R_{\rm Sch}$ is the Schwartzschild radius. Eq. (4) 
holds provided that $R_{5100}\gtrsim R_{\rm trap}$, or $\mathdotM\lesssim 
3\times 10^3m_7^{-1/2}$. No SEAMBH so far has exceeded this limit.}
\begin{equation}
\mathdotM=20.1\left(\frac{\ell_{44}}{\cos i}\right)^{3/2}m_7^{-2},
\end{equation}
where  $\ell_{44}$ is the 5100 \AA\, luminosity in units of $10^{44}\,\ergs$. 
This convenient expression can easily convert luminosity and BH mass into 
dimensionless accretion rates. In this paper, we take an average value of 
$\cos i = 0.75$, which corresponds to the opening angle of the dusty torus 
(e.g., \citealt{davis2011, du2015}). The uncertainties of $\mathdotM$ due to 
$i$ ($\in[0,45^{\circ}]$) are
$\Delta\log\mathdotM=1.5\Delta\log\cos i\lesssim 0.15$ from Equation (4),
where we took $\Delta \log \cos i\lesssim0.1$. This uncertainty is significantly
smaller than the average error bars of $\log \mathdotM$ ($\sim 0.35$), and is 
thus neglected.

The dimensionless accretion rate is related to the more widely used Eddington 
ratio via $\Lbol/L_{\rm Edd}=\eta\mathdotM$, where $\eta$ is the radiative 
efficiency, and $L_{\rm bol} \approx 10 L_{5100}$ \citep{kaspi2000}. 
The uncertainties of Eddington ratios result from the fact that the bolometric
correction depends on both accretion rates and BH mass \citep{jin2012}.  In 
our following discussion, we will use both $\mathdotM$ and $\Lbol/L_{\rm Edd}$.

\begin{figure*}[t!]
\begin{center}
\includegraphics[angle=0,width=0.82\textwidth]{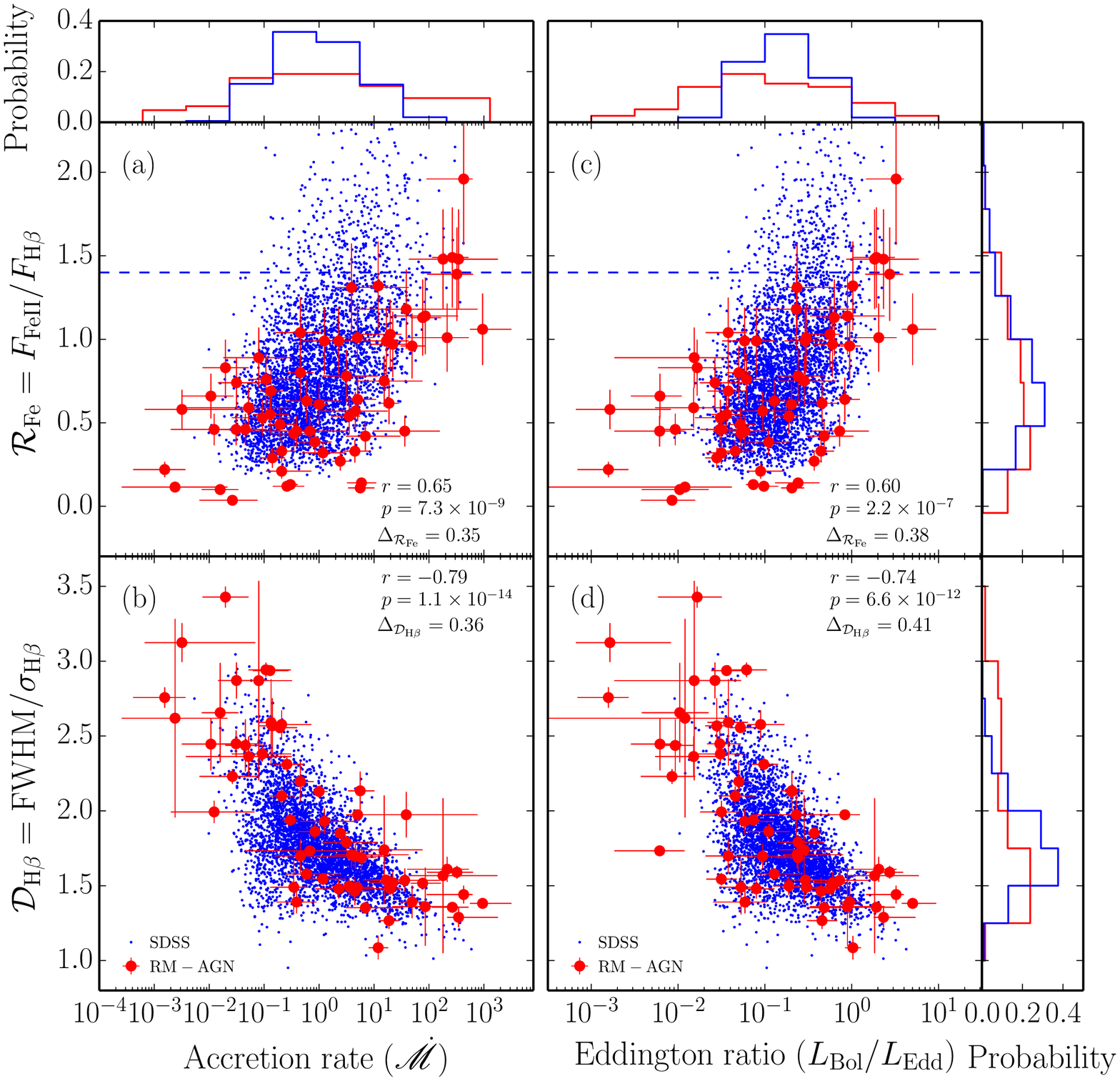}
\end{center}
\vglue -0.5cm
\caption{\footnotesize Correlations between ({\it a}) $\RFe-\mathdotM$ and 
({\it b}) $\Dhb-\mathdotM$.  The Pearson's coefficient, null-probability, 
and scatter of the $X-\mathdotM$ correlation are given by ($r,p,\Delta_X$). 
In panel ({\it a}), the SDSS quasars overlap with the RM AGNs quite well, 
except for AGNs with $\RFe\gtrsim 1.4$. This could be because these objects 
are super-Eddington accretors, in which the normal $R-L$ relation \citep{bentz2013} 
overestimates $\rhb$ as well as BH mass \citep{du2015}, and 
hence $\mathdotM$ is greatly underestimated (see details in the text). In 
panel ({\it b}), the SDSS sample also overlaps well with the RM AGNs, but the 
low$-\mathdotM$ AGNs lie beyond the locus of the SDSS sample. There are some 
SDSS quasars with extremely high accretion rates, $\mathdotM\gtrsim 10^2$, 
suggesting that we should monitor them in the future of SEAMBH project. The 
histograms indicate the distributions of $\RFe$, $\Dhb$ and $\mathdotM$ on 
a normalized scale. We note that there is no significant correlation between
$\RFe$ and $\Dhb$, either in the RM AGN or SDSS sample, indicating that 
$\Dhb$ and $\RFe$ are independent from each other, although both correlate 
with $\mathdotM$. Panels ({\it c}) and ({\it d}) are the same as ({\it a}) and 
({\it b}), but for Eddington ratios. 
}
\end{figure*}

\section{Fundamental plane of the BLR}
\subsection{Correlations}
Figure 1{\it a} and 1{\it c} show the $\RFe-(\mathdotM,L_{\rm bol}/L_{\rm Edd})$ 
plots and yield the following correlations:
\begin{equation}
\RFe=\left\{\begin{array}{l}
     (0.66\pm 0.04)+(0.30\pm 0.03)\log \mathdotM,\\[0.8em] 
     (1.20\pm 0.07)+(0.55\pm 0.06)\log \left(L_{\rm bol}/L_{\rm Edd}\right).
     \end{array}\right.
\end{equation}
We define the scatter of a correlation as 
$\Delta_{X}=\sqrt{\sum_{i=1}^N(X-X_i)^2/N}$, where $N$ is the number of 
objects, and $X$ represents $\RFe$, $\Dhb$, $\mathdotM$, or $\Lbol/\LEdd$.  
The Pearson's correlation coefficient ($r$), null-probability ($p$), and 
scatters are indicated in the plots. By comparing $(r,p,\Delta_{\rm \RFe})$ in 
panels ({\it a}) and ({\it c}), we find that the $\RFe-\mathdotM$ correlation 
is slightly stronger than that of $\RFe-L_{\rm bol}/L_{\rm Edd}$.  In 
high-$\mathdotM$ AGNs, both H$\beta$ and continuum variability are 
significantly smaller than those in sub-Eddington AGNs. On the other hand, 
\feii\, reverberates in a very similarly fashion to H$\beta$ with respect to 
the continuum \citep{hu2015}. Indeed, it can be seen that the scatter of the 
correlation gets larger with decreasing $\mathdotM$ or 
$L_{\rm bol}/L_{\rm Edd}$. The $\RFe-(\mathdotM, \Lbol/\LEdd)$ correlations 
supports the idea that \feii\, strength is not governed by metallicity 
but by the ionizing flux and hydrogen density \citep{verner2004}. 

We plot the $\Dhb-(\mathdotM, \Lbol/\LEdd)$ relations in Figure 1{\it b} and 
1{\it d} and find
\begin{equation}
\Dhb =\left\{\begin{array}{l}
           (2.01\pm 0.05)-(0.39\pm 0.04)\log \mathdotM,\\[0.8em]
            (1.28\pm0.09)-(0.72\pm0.08)\log\left(L_{\rm bol}/L_{\rm Edd}\right).
           \end{array}\right.
\end{equation}                   
The above two correlations are similar, but the former is slightly stronger 
than the latter. \cite{collin2006} also found a correlation between 
$\Dhb$ and $\Lbol/\LEdd$ (see their Figure 6), but their results are much 
weaker than ours. This is mainly due to the lack of high$-\mathdotM$ AGNs in
their sample. We would like to emphasize that the 
$\Dhb-(\mathdotM,\Lbol/\LEdd)$ correlations cannot be an artifact of the 
inclusion of FWHM in $\mathdotM$.  For a constant $\sighb$ of RM AGNs, the 
accretion rates span over about 5 dex whereas luminosities span over 4.5;
however, the $\Dhb-\mathdotM$ relation has a scatter of only
$\Delta_{_{\cal D}}=0.3-0.4$. The correlations are intrinsic. 

Figure 1 also shows, as background, the SDSS DR5 sample of \cite{hu2008}. 
The sample comprises 4037 $z\lesssim 0.8$ 
quasars with criteria of S/N$\ge 10$ and EW(\feii)$\ge 25$ \AA\, (this excludes \feii-weak 
quasars). BH masses assume $f_{\rm BLR} = 1$ and a standard $R-L$ relation\footnote{ 
This is an empirical relation between the BLR size and the continuum.  From the recent work 
of Bentz et al. (2013), it has the form  $R_{\rm BLR}=33.65\,\ell_{44}^{0.53}$ltd. However, 
\cite{du2015} found that it only applies to sub-Eddington AGNs; it depends on $\mathdotM$ 
for super-Eddington AGNs. We do not consider 
the dependence of the $R-L$ relation on $\mathdotM$ for the SDSS sample in this paper.}.  
The RM AGNs overlap very well with the SDSS sample, on both the 
$\RFe-(\mathdotM,\Lbol/\LEdd)$ and the $\Dhb-(\mathdotM,\Lbol/\LEdd)$ plots. 
We note that among the mapped AGNs there is a small population ($\lesssim 9\%$; Figure 
1{\it a} and {\it c}) of AGNs with $\RFe>1.4$ of what appear to be super-Eddington sources.  
Their values of $\mathdotM$ are likely underestimated because their black hole masses were 
estimated using the standard $R-L$ relation.

\begin{figure*}[!t]
\begin{center}
\includegraphics[angle=0,width=0.45\textwidth]{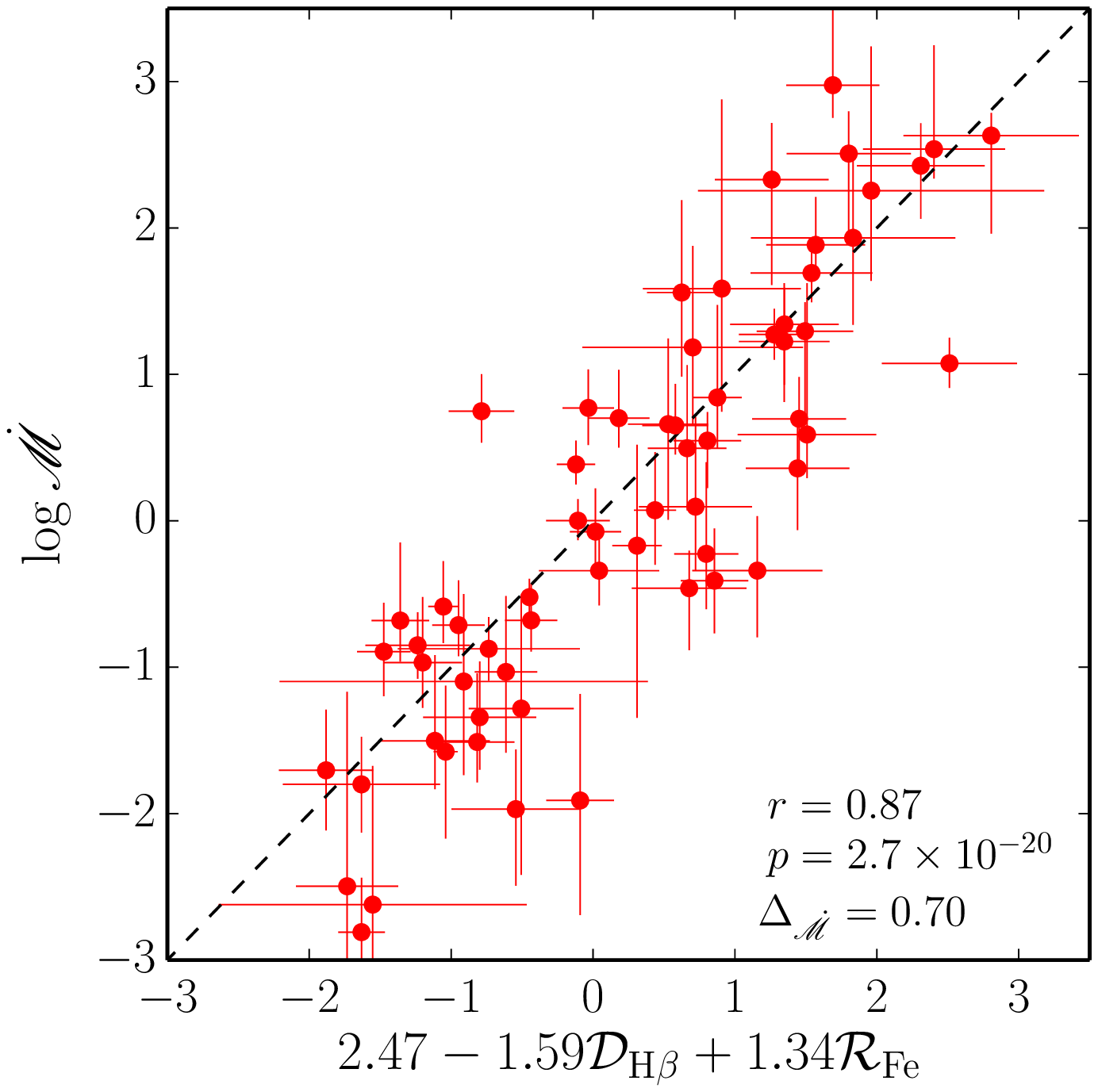}
\includegraphics[angle=0,width=0.45\textwidth]{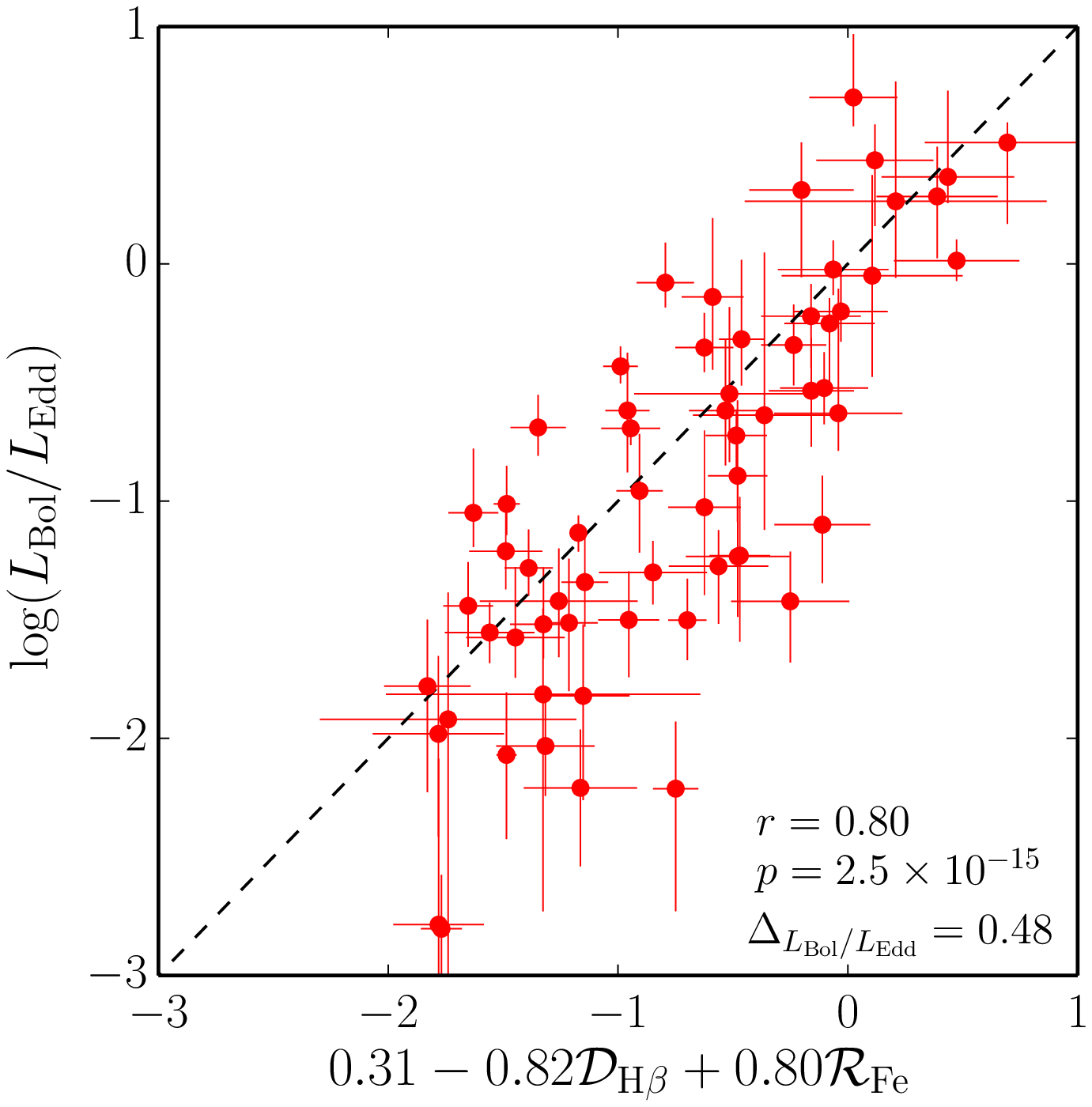}
\end{center}
\vglue -0.5cm
\caption{\footnotesize The fundamental plane of AGN BLRs, showing a physical 
connection between accretion disks and BLRs. 
The dependent variable is ({\it left}) $\mathdotM$ and ({\it right}) 
Eddington ratio. The two observables of $\RFe$ and $\Dhb$ can be readily 
measured from single-epoch spectra, allowing us to constrain the accretion 
status of the central engine. }
\end{figure*}

\subsection{Fundamental Plane}
The $(\RFe,\Dhb)-(\mathdotM,\Lbol/\LEdd)$ relations reflect 
connections between the BLR structure and dynamics with BH accretion.  We 
investigate whether these two univariate correlations can be unified into a 
single bivariate correlation of the form
\begin{equation}
\log (\mathdotM, L_{\rm bol}/L_{\rm Edd})=\alpha_k + \beta_k\Dhb+\gamma_k\RFe,
\end{equation}
where $(\alpha_k,\beta_k,\gamma_k)$ are coefficients to be determined by 
data ($k=1,2$).  We define
\begin{equation}
\chi_k^2=\frac{1}{N}\sum_{i=1}^N
       \frac{\left(\log \calA_k^i-\alpha_k-\beta_k\Dhb^i-\gamma_k\RFe^i\right)^2}
            {\sigma_{\scriptscriptstyle{\calA_i}}^2+
            \beta_k\sigma_{\scriptscriptstyle{\Dhb^i}}^2+\gamma_k\sigma_{\scriptscriptstyle{\RFe^i}}^2},
\end{equation}
where $\calA_k=(\mathdotM,L_{\rm bol}/L_{\rm Edd})$, 
$\sigma_{\scriptscriptstyle{\calA_i}},\sigma_{\scriptscriptstyle{\Dhb^i}}$ 
and $\sigma_{\scriptscriptstyle{\RFe^i}}$ are the error bars of $\log\calA$, 
$\Dhb$, and $\RFe$ of the $i-$th object, respectively. Minimizing $\chi_k^2$,
we obtain 
$$
\alpha_1=2.47\pm 0.34;~~~\beta_1=-1.59\pm 0.14;~~~{\rm and}~~\gamma_1=1.34\pm 0.20,
$$
$$
\alpha_2=0.31\pm 0.30;~~~\beta_2=-0.82\pm 0.11;~~~{\rm and}~~\gamma_2=0.80\pm 0.20.
$$
The error bars of $(\alpha_k,\beta_k,\gamma_k)$ are derived from bootstrap 
simulations.  The bivariate correlations, plotted in Figure 2, are much 
stronger than individual corrections of Figure 1 (see the correlation 
coefficients and null-probability).  We call these new correlations as the 
fundamental plane of the BLR.

The implications of Equation (7) are exciting.   From two simple measurements 
of a single-epoch spectrum of a quasar---strength of \feii\ and shape of broad 
H$\beta$---we can deduce the status of its accretion flow.  This can be very 
useful when applied to large samples of quasars to investigate the cosmological 
growth of BHs.  Our method can be usefully applied to quasars with suitable 
spectroscopy in the rest-frame H$\beta$ region, for which the strength of \feii\ 
can be measured or constrained.

\subsection{Application to SDSS sample}
We apply the $\mathdotM-$plane (Equation 7) to a sample of 
4037 objects \cite{hu2008}, which were selected from the
SDSS DR5 sample composed of $N_{\rm tot}\approx 15,000$ quasars with $z\lesssim 0.8$.  
We calculate fractions of quasars with $\mathdotM\ge \mathdotM_c$,
$\delta=N_{{\mathdotM_c}}/N_{\rm tot}$, where $N_{{\mathdotM_c}}$ 
is the number of quasars and
$\mathdotM_c$ is the critical accretion rate in question. For objects with $\mathdotM\ge 3$, we 
find $\delta_3=N_3/N_{\rm tot}\approx 0.18$. Similarly,
we have $\delta_{10}=N_{10}/N_{\rm tot}\approx 0.12$ and 
$\delta_{100}=N_{100}/N_{\rm tot}\approx 0.02$.
These numbers show that
super-Eddington accreting AGNs are quite common in the Universe at $z<0.8$. We should note 
that these fractions are lower limits, as a result of the selection criteria imposed by 
Hu et al.  Detailed results of the application of our technique to the latest sample of SDSS 
quasars will be carried out in a separate paper.

\section{Conclusions}
This paper studies correlations among three dimensionless AGN parameters: 
accretion rate (or Eddington ratio), shape of the broad H$\beta$ line, and 
flux ratio of optical \feii\, to H$\beta$.  A strong correlation among them is 
found, which we denote as the fundamental plane of AGN BLRs (Equation 7). 
The BLR fundamental plane enables us to conveniently explore the accretion 
status of the AGN central engine using single-epoch spectra, opening up many 
interesting avenues for exploring AGNs, including their cosmological evolution.
A simple application of the BLR fundamental plane shows that super-Eddington 
accreting AGNs are quite common in among low-redshift quasars.

\acknowledgements{ This research 
is supported by the Strategic Priority Research Program - The Emergence of Cosmological Structures 
of the Chinese Academy of Sciences, Grant No. XDB09000000, by NSFC grants NSFC-11173023, -11133006, 
-11373024, -11233003 and -11473002, and a NSFC-CAS joint key grant of U1431228.}

\end{document}